\documentclass[twoside,twocolumn,british,aps,prl,superscriptaddress]{revtex4}
\usepackage[T1]{fontenc}
\usepackage[latin1]{inputenc}
\usepackage{amsmath}

\usepackage{graphicx}
\makeatletter

\usepackage{babel}
\makeatother
\begin{document}
\newcommand \beq{\begin{equation}}
\newcommand \beqa{\begin{eqnarray}}
\newcommand \beqann{\begin{eqnarray*}}
\newcommand \eeq{\end{equation}}
\newcommand \eeqa{\end{eqnarray}}
\newcommand \eeqann{\end{eqnarray*}}
\newcommand \ga{\raisebox{-.5ex}{$\stackrel{>}{\sim}$}}
\newcommand \la{\raisebox{-.5ex}{$\stackrel{<}{\sim}$}}

\title{\textmd{Exact Results for Three-Body Correlations in a Degenerate
One-Dimensional Bose Gas }}

\author{Vadim V. Cheianov}
\affiliation{NORDITA, Blegdamsvej 17, DK-2100 Copenhagen, Denmark}
\author{H. Smith}
\author{M. B. Zvonarev}
\affiliation{Niels Bohr Institute, Universitetsparken 5, DK-2100
Copenhagen, Denmark}

\begin{abstract}
Motivated by  recent experiments we derive an exact expression for
the correlation function entering  the three-body recombination
rate for a one-dimensional  gas of interacting bosons. The answer,
given in  terms of two thermodynamic parameters of the
Lieb-Liniger model, is valid for all values of the dimensionless
coupling $\gamma$ and contains the previously known results for
the Bogoliubov and Tonks-Girardeau regimes as limiting cases. We
also investigate finite-size effects by calculating the
correlation function for small systems of 3, 4,  5 and 6
particles.
\end{abstract}
\maketitle Quantum fluctuations are well known to have a profound
influence on the physics of one-dimensional (1D) systems. Due to
quantum fluctuations, a 1D gas of weakly interacting bosons does not
exhibit  true long-range order and hence does not undergo
Bose-Einstein condensation at any temperature. With increasing
repulsion between the bosons, the gas exhibits a smooth crossover to
the strong coupling regime, where the  correlations in the positions
of the particles endow it with fermion-like thermodynamic
properties. A strong repulsion between the bosons also has a marked
effect on the local fluctuation properties, leading, for example, to
a fermion-like suppression of the fluctuations of particle density.
In this paper we shall focus on a three-body correlation function,
which is directly related to the lifetime of a Bose gas unstable
with respect to three-body recombination processes. It has been
studied in a recent experiment, where measurements of the three-body
recombination rate for trapped atoms were performed~\cite{Tolra}.

The idea  of using three-body recombination for a measurement of
local correlations was originally put forward by Kagan et
al.~\cite{KSS-85}. These authors showed that the low-temperature
recombination rate for a 3D Bose gas is proportional to the local
three-body correlation function $g_3$ defined by the ground-state
expectation value
\begin{equation}
g_3=\langle :\hat{n}^3(\mathbf r):\rangle \equiv
\langle\hat{\Psi}^{\dagger}({\bf r})^3\hat{\Psi}({\bf
r})^3\rangle.\label{3corr}
\end{equation}
Here $\hat{n}(\mathbf r)=\hat{\Psi}^{\dagger}(\mathbf r)
\hat{\Psi}(\mathbf r) $ is the particle density operator,
$\hat{\Psi}^{\dagger}$ ($\hat{\Psi}$) denote boson creation
(annihilation) operators and the symbol $::$ stands for normal
ordering. It was noted that, due to the factor-of-six suppression of
$g_3$ in a Bose-Einstein condensate at zero temperature as compared
to the non-condensed Bose gas, three-body recombination can be used
as a diagnostic tool for distinguishing between the condensed and
the non-condensed phases. This was later confirmed
experimentally~\cite{Burt}.

In the experiment of Ref.~\cite{Tolra} the recombination process was
used for the investigation of three-body correlations in a 1D Bose
gas. A magnetically trapped Bose-Einstein condensate of $^{87}$Rb
atoms was loaded into a deep 2D optical lattice, by which the
condensate was divided into an array of independent 1D systems. In
the case of $^{87}$Rb two-body losses are very small, and it was
possible to model the decay in time of the number of trapped atoms
in terms of one-body and three-body processes only. The rate of
change of the total number of atoms $N$ was written as
\begin{equation}
\frac{dN}{dt}=-K_1N-\int d\mathbf{r}\, K_3^{\rm 1D}n_{\rm
3D}^3,\label{rate1}
\end{equation}
where $K_1$ and $K_3^{\rm 1D}$ are the rate coefficients for one-
and three-body losses. By measuring the decrease of the number of
atoms as a function of time it was found that the rate coefficient
$K_3^{\rm 1D}$ in the 1D gas was reduced considerably, by a factor
of about 7, compared to its value $K_3^{\rm 3D}$ for a 3D
condensate. The measurements were carried out for a particular value
of the dimensionless coupling constant $\gamma$ ($\approx 0.5$)
characterizing the strength of the 1D correlations. Given that $K_3$
is proportional to the three-body correlation function $g_3$, the
observed reduction was interpreted as an effect of the reduced
dimensionality on~$g_3$.

In the following  we shall therefore focus on calculating $g_3$
for a 1D Bose gas. Our treatment is based on the Lieb-Liniger (LL)
model \cite{Lieb} for which  the dimensionless coupling parameter
$\gamma$ is given in Eq.~\eqref{dimcoupl} below. In the
thermodynamic limit, where the particle number $N$ and the size of
the system $L$ tend to infinity while the density $n$ remains
constant,
\begin{equation}
n=N/L=\mathrm{const}, \quad N,L\to\infty, \label{tl}
\end{equation}
an expression for $g_3$ has so far been obtained only for small and
large values of $\gamma$ \cite{Gangardt}. Here we report an explicit
expression for $g_3,$ valid for all $\gamma.$ Its derivation employs
an integrable lattice regularization of the LL
model~\cite{BBP-93,BIK-98} together with conformal field theory. The
calculations are quite involved and will be presented
elsewhere~\cite{CSZ}. The answer is given by Eq.~\eqref{g3result} in
terms of two thermodynamic parameters of the LL model: the second-
and fourth-order moment, Eq.~\eqref{b3}, of the quasi-momentum
distribution function. The quasi-momentum distribution function is
the solution of the Lieb equation \eqref{b1}, which is a linear
integral equation.

Within the LL model the Hamiltonian for $N$ identical bosons of mass
$m$ is taken to be
\begin{equation}
H=\frac{\hbar^2}{2m}\left[-\sum_{i=1}^N\frac{\partial^2}{\partial
x_i^2}+2c\sum_{1\le i<j\le N}\delta(x_i-x_j)\right]. \label{LL2}
\end{equation}
The interaction constant $c\ge0$ has the dimension of inverse
length. The dimensionless coupling strength $\gamma$ is given by
\begin{equation}
\gamma=c/n.\label{dimcoupl}
\end{equation}
The LL model allows for an exact determination of the eigenfunctions
and eigenvalues of $H$. In the $\gamma\rightarrow\infty$ limit of
the model, known as the Tonks-Girardeau (TG) gas~\cite{Tonks}, the
eigenvalues are the same as that of a free fermion gas.

While the eigenfunctions and eigenvalues are known exactly for the
LL model, its correlation functions in general and $g_3$ in
particular have been much less investigated. The reason for this is
that the eigenfunctions are very complicated for a general value of
$N$.
We exhibit these below and
calculate $g_3$ analytically for the simplest possible  case
$N=3.$

The eigenfunctions $\Psi(x_1,x_2,\ldots,x_N)$ of the Hamiltonian
\eqref{LL2} must be symmetric under any interchange of coordinates.
Since the particles interact via a delta-function potential, the
derivatives of $\Psi$ jumps when two particles approach each other,
while when all $x_j$ are different, $\Psi$ satisfies the
free-particle Schr\"odinger equation. The explicit expression for
$\Psi$ was obtained in Ref.~\cite{Lieb} and given in
Ref.~\cite{KBI-93} in the following form
\begin{multline}
\Psi(x_1,x_2,\ldots,x_N)=
C\sum_P(-1)^{[P]}\exp\left\{i\sum_{j=1}^N k_{P_j}x_j\right\}\\
\times\prod_{j>l}[k_{P_j}-k_{P_l}-ic\mathrm{sgn}(x_j-x_l)],
\label{fB}
\end{multline}
where $P$ is a permutation of $N$ numbers and $[P]$ is the parity
of the permutation. The possible values of the quasi-momenta $k_j$
are determined by the boundary conditions imposed on the system.
For  periodic boundary conditions (that is, for $N$ particles
placed on a ring of circumference $L$) $k_j$ are solutions to the
following system of coupled nonlinear equations, called the Bethe
equations,
\begin{equation}
\exp\left\{ik_jL\right\}=\prod_{l\ne j}^N
\frac{k_j-k_l+ic}{k_j-k_l-ic}, \qquad j=1,\dots,N. \label{Bethe}
\end{equation}
The calculation of the normalization constant $C$ in
Eq.~(\ref{fB}) is a nontrivial task. For the periodic boundary
conditions a closed expression for $C$ was suggested by Gaudin
(see~\cite{Gaudin-83} and references therein) and proved in
Ref.~\cite{Korepin-82}; a detailed discussion can be found in
Ref.~\cite{KBI-93}. One has now all the ingredients for
calculating $g_3\equiv g_3(\gamma,N)$ for the 1D system. Written
in the first-quantized form, $g_3(\gamma,N)$ is
\begin{equation}
g_3(\gamma,N)=\frac{N!}{(N-3)!}\int_0^L
dX\,|\Psi(0,0,0,x_4,\ldots,x_N)|^2, \label{g3}
\end{equation}
where $dX=dx_4\dots dx_N$ and $\Psi$ is given by Eq.~\eqref{fB}.
Note that the average $\langle\dots\rangle$ in Eq.~\eqref{3corr}
is taken over the ground state of the system; equivalently, it is
the ground-state wave function which enters Eq.~\eqref{g3}. This
is ensured by selecting the proper set of the quasi-momenta $k_j$
among all the solutions of the Bethe equations \eqref{Bethe}.

All  quasi-momenta $k_j$ are equal to zero for the non-interacting
system, $\gamma=0,$ and the wave function \eqref{fB} is uniform in
space, $\Psi=(1/\sqrt{L})^N$ , which implies that
\begin{equation}
g_3(0,N)=n^3{N(N-1)(N-2)}/{N^3}.
\end{equation}

\begin{figure}
\begin{center}
\includegraphics[clip,width=8cm]{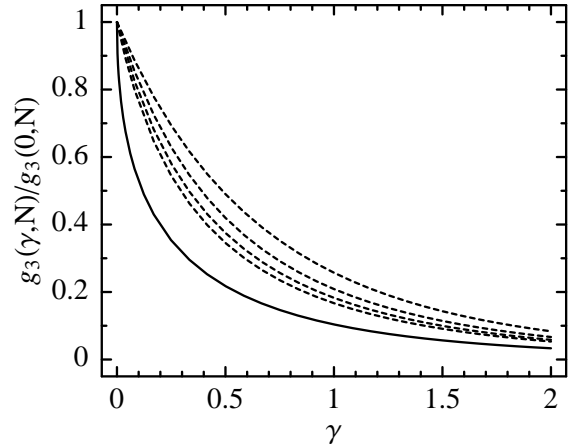}
\end{center}
\caption{The three-body correlation function $g_3$, normalized to
its value in the absence of interactions, as a function of
$\gamma$. The dashed lines are for particle numbers $N=3,4,5,$ and
$6$ (from top to bottom), and the full line is for $N=\infty$.}
\label{g3plot}
\end{figure}

We now use the wave function \eqref{fB} to get $g_3(\gamma,N)$ for
$N=3$ analytically, and for $N=4,5$ and $6$ numerically. The
quasi-momenta in the ground state of the system with $N=3$ obey
$k_1=-k_3$ and $k_2=0.$ It is convenient to write the Bethe equation
for $k_1$ in the form
\begin{equation}
\lambda=2\pi-2\arctan(\lambda/3\gamma)-
2\arctan(2\lambda/3\gamma),
\end{equation}
where we have introduced $\lambda\equiv k_1L$ and the principal
branch for arctangent has been chosen:
$-\pi/2\le\arctan(x)\le\pi/2.$ The solution to this transcendental
equation grows monotonically from $\lambda\simeq 3\sqrt{\gamma}$
for $\gamma\ll 1$ to $\lambda\simeq 2\pi-4\pi/\gamma$ for
$\gamma\gg 1$. The resulting answer for
$g_3(\gamma,3)=6|\Psi(0,0,0)|^2$ is
\begin{equation}
g_3(\gamma,3)= \frac{16}3\frac{A^2B\lambda^6}{\gamma^3
L^3[1+4(3A^2+2A+B+6AB)]},\label{g3gamma3}
\end{equation}
where $A=3\gamma/(9\gamma^2+\lambda^2)$ and
$B=3\gamma/(9\gamma^2+4\lambda^2).$ The ratio
$g_3(\gamma,3)/g_3(0,3)$ is plotted in Fig.~\ref{g3plot}.
It decreases monotonously with increasing $\gamma,$ from
$1-{29}\gamma/24$
for $ \gamma\ll 1$ to $512 \pi^6/243\gamma^6$
for $ \gamma\gg 1$. In Fig.~\ref{g3plot} we also plot
$g_3(\gamma,N)$ calculated for $N=4$, $5$ and $6.$ Note that the
magnitude of the slope at the origin ($\gamma$=0) increases with
increasing number of particles, and when $N\rightarrow\infty$ it
approaches infinity, in agreement with the Bogoliubov result given
in Ref.~\cite{Gangardt}:
\begin{equation}g_3(\gamma,
\infty)/n^3\simeq 1-6\sqrt{\gamma}/\pi, \quad
\gamma\to0.\label{Misha}
\end{equation}
We have obtained an analytic expression for $g_3$ in the
Tonks-Girardeau limit, $\gamma\to\infty,$ for arbitrary values of
$N$ by generalizing the result Eq.~(16) of  Ref.~\cite{Gangardt},
valid in the thermodynamic limit, to the case of a finite number of
particles. The result is
\begin{equation}
\frac{g_3(\gamma,N)}{n^3}\simeq\frac{16\pi^6}{15\gamma^6}\frac{(N^2-1)^2(N^2-4)}{N^6},
\quad \gamma\to\infty.\label{g3largegamma}
\end{equation}

We now present our main result: the exact expression for the
three-body correlation function $g_3(\gamma,\infty)$ in the
thermodynamic limit \eqref{tl} of the Lieb-Liniger model
\eqref{LL2}. The system of Bethe equations \eqref{Bethe} reduces in
this limit to the linear integral equation called the Lieb equation:
\begin{equation}
\sigma(k)-\frac1{2\pi}\int_{-1}^1dq\,\frac{2\alpha\sigma(q)}
{\alpha^2+(k-q)^2}=\frac1{2\pi}, \label{b1}
\end{equation}
where $\alpha$ is an implicit function of $\gamma:$
\begin{equation}
\alpha=\gamma\int_{-1}^1dk\,\sigma(k) \label{b2}.
\end{equation}
In terms of the moments $\epsilon_m$ defined by
\begin{equation}
\epsilon_m=\left(\frac{\gamma}{\alpha}\right)^{m+1}
\int_{-1}^{1}dk\,k^m\sigma(k), \quad m=2,4.\label{b3}
\end{equation}
the exact expression for $g_3(\gamma,\infty)$ is
\begin{multline}
\frac{g_3(\gamma,\infty)}{n^3}=
\frac{3}{2\gamma}\epsilon_4'-\frac{5\epsilon_4}{\gamma^2}
\\+\left(1+\frac{\gamma}{2}\right)\epsilon_2'-2\frac{\epsilon_2}{\gamma}-
\frac{3\epsilon_2\epsilon_2'}{\gamma}+\frac{9\epsilon_2^2}{\gamma^2},
\label{g3result}
\end{multline}
where $\epsilon^\prime_m$ is the derivative of $\epsilon_m$ with
respect to $\gamma.$

\begin{figure}
\begin{center}
\includegraphics[clip,width=8cm]{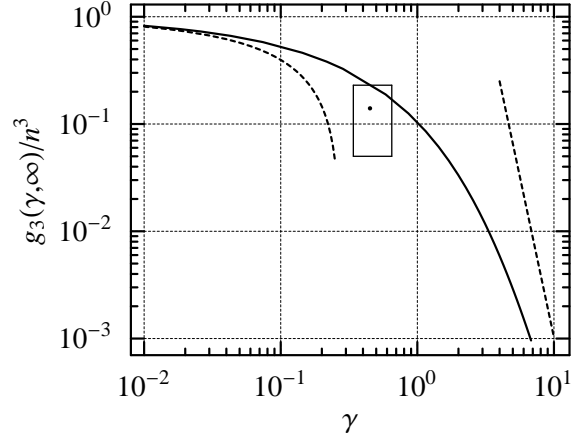}
\end{center}
\caption{Double-logarithmic plot of the three-body correlation
function $g_3(\gamma,\infty)$ as a function of $\gamma$ (full
line). The dashed lines indicate the asymptotic
results~\cite{Gangardt} given in Eqs.~\eqref{Misha} and
\eqref{g3largegamma}  for small and large $\gamma$, respectively.
The measured value is shown as the dot inside a box indicating the
experimental uncertainty~\cite{Tolra}.} \label{g3logplot}
\end{figure}

Equations \eqref{b1}--\eqref{g3result} form a closed set of
equations determining $g_3(\gamma,\infty).$ We plot the result in
Figs.~\ref{g3plot} and \ref{g3logplot}.
For a numerical solution of the integral equation \eqref{b1} we have
used the \textsf{Mathematica} package \textsf{NISolve}
\cite{NISolve}. In Fig.~\ref{g3plot} we compare the normalized
function $g_3(\gamma,\infty)/g_3(0,\infty)$ with that calculated for
$N=3$, $4$, $5$, and $6.$ One can see that the convergence to the
thermodynamic limit is quite slow. In Fig.~\ref{g3logplot} we
compare $g_3(\gamma,\infty)/n^3$ with the experimental data from
Ref.~\cite{Tolra}. The average value of $\gamma,$ measured in
\cite{Tolra}, $\gamma_{\rm m}\approx0.45,$ and the corresponding
value of $g_3(\gamma_{\rm m},\infty)\approx0.14$ are shown as the
dot inside the box which represents the experimental uncertainty
according to Ref.~\cite{Tolra}: $0.34<\gamma_{\rm m}<0.65$ and
$0.05<g_3(\gamma_{\rm m},\infty)<0.23.$ The dashed lines show the
asymptotic expressions given in Eqs.~\eqref{Misha} and
\eqref{g3largegamma} for small and large $\gamma,$ respectively.
Evidently,  the asymptotic expressions do not account for the
observed value, and the exact expression \eqref{g3result} is needed.

Eq.~\eqref{g3result} gives $g_3(\gamma,\infty)/n^3\simeq
16\pi^6/15\gamma^6$ when $\gamma\to\infty,$ thus reproducing the
asymptotic expression Eq.~\eqref{g3largegamma} taken at $N=\infty.$
To check this is a straightforward task since Eq.~\eqref{b1} admits
a regular perturbative expansion with $1/\alpha$ being a small
parameter. The opposite limit, $\gamma\to0,$ of Eq.~\eqref{g3result}
is much more difficult to analyze since the kernel in Eq.~\eqref{b1}
becomes singular when $\alpha\to0.$ Beyond the leading order, the
results were obtained only recently \cite{Wadati}:
\begin{equation}
\sigma(k)\simeq\sqrt{1-k^2}/2\pi\alpha+f(k), \quad \alpha\to0
\label{sigma}
\end{equation}
where $f(k)$ is written explicitly in Ref.~\cite{Wadati}, see
Eq.~(4.11). For the moments Eq.~\eqref{b3} one gets
$\epsilon_2\simeq\gamma(1-4\sqrt{\gamma}/3\pi)$ and
$\epsilon_4\simeq2\gamma^2(1-44\sqrt\gamma/15\pi).$ With these
expressions for the moments, expansion \eqref{Misha} is reproduced.


A useful representation of our result (\ref{g3result}) is the
following approximate form
\begin{align}
&\frac{1-6\pi^{-1}\gamma^{\frac12}+1.2656\gamma-0.2959\gamma^{\frac32}}
{1-0.2262\gamma-0.1981\gamma^{\frac32}},&& 0\le\gamma\le 1,\label{intpol1}\\
&\frac{0.705-0.107\gamma+5.08\cdot10^{-3}\gamma^2}
{1+3.41\gamma+0.903\gamma^2+0.495\gamma^3}, && 1\le\gamma\le 7,\\
&\frac{16\pi^6}{15\gamma^6}\frac{9.43-5.40\gamma+\gamma^2}
{89.32+10.19\gamma+\gamma^2}, && 7\le\gamma\le30.\label{intpol3}
\end{align}
The relative error of this approximation does not exceed
$2\cdot10^{-3}$ for all values of $\gamma$ in the interval $0\le
\gamma\le 30$, which we expect to be experimentally relevant. Note
that the asymptotic result (\ref{g3largegamma}) for $N=\infty$ is
nearly a factor of two larger than the exact result for
$\gamma=30$.

The formal derivation of Eq.~\eqref{g3result} is quite lengthy and
will be given elsewhere~\cite{CSZ}. Here we shall only summarize
the main steps: We are interested in the thermodynamic limit of
the model, defined by Eq.~\eqref{tl}. In this limit it is natural
to use a field-theoretical approach rather than the
first-quantized formalism we employed for the few-particle
systems. The Lieb-Liniger model is completely integrable. In the
field-theoretical language complete integrability  implies the
existence of an infinite family of conserved currents
$\mathcal{J}_n,$ $n=0,1,2,\ldots$. The integrals over space of
these conserved currents commute with each other, including the
Hamiltonian \eqref{LL2}. There is thus an infinite set of mutually
commuting operators $J_n$, $n=0,1,2,\ldots,$ for which the
eigenfunctions and the spectrum are known exactly. A procedure for
generating $J_n$ is discussed, for example, in Ref.~\cite{KBI-93}.
The first three currents generated by this method are
$\mathcal{J}_0=\mathcal{N},$ $\mathcal{J}_1=\mathcal{P},$ and
$\mathcal{J}_2=\mathcal{H},$ where $\mathcal{N},$ $\mathcal{P},$
and $\mathcal{H}$ are the number density of particles, the
momentum density, and the Hamiltonian density, respectively. These
three currents exist for non-integrable models as well, while the
currents $\mathcal{J}_n$ for $n>2$ are specific to the LL model.
To illustrate our approach we introduce the classical expression
for the current $\mathcal{J}_4$ (its quantization will be
discussed in the next paragraph):
\begin{multline}\mathcal{J}_4=
\Psi^*\partial_x^4\Psi + c \Psi^*\Psi^* (\partial_x\Psi)\partial_x
\Psi + 8 c (\partial_x\Psi^*)  \Psi^* \Psi \partial_x\Psi \\ +c
(\partial_x\Psi^*) (\partial_x\Psi^*) \Psi \Psi +2 c^2  (\Psi^3)^*
\Psi ^3. \label{j3t}
\end{multline}
The quantum-mechanical version of the last term in Eq.\
(\ref{j3t}) contains $\hat\Psi^\dagger(x)^3\hat\Psi(x)^3$ in which
we are interested. When averaged over the ground state, this term
gives the desired correlation function $g_3,$ Eq.~\eqref{3corr}.
Some of the remaining terms  in (\ref{j3t}) can be eliminated by
using the Hellmann-Feynman theorem, which applies to all conserved
quantities. This theorem was applied to the Hamiltonian of the LL
model in Ref.~\cite{Gangardt} to calculate the correlation
function $\langle\hat\Psi^\dagger(x)^2\hat\Psi(x)^2\rangle$. In
our case not all terms can be eliminated by use of the
Hellmann-Feynman theorem. We have succeeded in obtaining
additional identities~\cite{CSZ} by employing the long-distance
asymptotics of the correlation functions of conserved currents
known from the conformal limit of the theory.

The quantization of the classical current Eq.~(\ref{j3t}) is far
from being straightforward. A naive quantum-mechanical version of
the expression (\ref{j3t}) contains terms involving derivatives of
field operators $\hat{\Psi}$ and $\hat{\Psi}^{\dagger}$. Such
terms are not  well-defined in general. An explicit demonstration
of this was considered in the paper~\cite{Ol}, where it was shown
that the expectation value $\langle
\hat\Psi^\dagger(0)\partial^3_x\hat\Psi(x)\rangle$ is
discontinuous at $x=0$. The reason for this behavior can be seen
from the structure of the Bethe-ansatz wave function \eqref{fB},
which has a cusp whenever the coordinates of any two particles
coincide. This does not cause difficulties when one considers the
momentum operator and the Hamiltonian itself, that is, when one
works with the operators
$\hat\Psi^\dagger(x)\partial_x\hat\Psi(x)$ and
$\hat\Psi^\dagger(x)\partial^2_x\hat\Psi(x).$ However, for other
operators  we encounter difficulties which we have not been able
to resolve within the continuum model \eqref{LL2}. Therefore we
have used an integrable lattice model, the so-called $q$-boson
hopping model \cite{BBP-93,BIK-98}, which reduces to the
Lieb-Liniger model \eqref{LL2} in the continuum limit. Within this
model all  higher (quantum) integrals of motion are defined
unambiguously.

In conclusion, we have obtained an exact expression for the
three-body correlation function Eq.~\eqref{3corr} within the
Lieb-Liniger model of a one-dimensional Bose gas. The explicit
expression is given by Eq.~\eqref{g3result} and Eqs.\
(\ref{intpol1})-(\ref{intpol3}), and compared in
Fig.~\ref{g3logplot} with a recent experiment. The finite-size
effects are explored and a comparison with the thermodynamic limit
is given in Fig.~\ref{g3plot}. We expect that our results will
stimulate further experimental investigations of the local
correlations in low-dimensional trapped atomic gases.

The authors would like to thank N.M.~Bogoliubov for helpful
discussions. M.B.~Zvonarev's work was supported by the Danish
Technical Research Council via the Framework Programme on
Superconductivity.


\begin{thebibliography} {99}
\bibitem{Tolra}
B.~Laburthe Tolra, K.M.~O'Hara, J.H.~Huckans, W.D.~Phillips,
S.L.~Rolston, and J.V.~Porto, Phys. Rev. Lett. {\bf 92}, 190401
(2004).
\bibitem{KSS-85} Y.~Kagan, B.V.~Svistunov, and G.V.~Shlyapnikov,
JETP Lett. {\bf 42}, 209 (1985).
\bibitem{Burt} E.A.~Burt, R.W.~Ghrist, C.J.~Myatt, M.J.~Holland, E.A.~Cornell, and
C.E.~Wieman, Phys. Rev. Lett. {\bf 79}, 337 (1997).
\bibitem{Lieb}E.H.~Lieb and W.~Liniger,
Phys. Rev. {\bf 130}, 1605 (1963).
\bibitem{Gangardt} D.M.~Gangardt and G.V.~Shlyapnikov, Phys. Rev. Lett. {\bf 90}, 010401 (2003).
\bibitem{BBP-93} N.M.~Bogoliubov, R.K.~Bullough, and G.D.~Pang,
Phys. Rev. B {\bf 47}, 11495 (1993).
\bibitem{BIK-98} N.M.~Bogoliubov, A.G.~Izergin, and N.A.~Kitanine,
Nucl. Phys. B {\bf 516}, 501 (1998).
\bibitem{CSZ} Vadim V.~Cheianov, H.~Smith, and M.B.~Zvonarev,
unpublished.
\bibitem{Tonks} L.~Tonks, Phys. Rev. {\bf 50}, 955 (1936); M.~Girardeau, J. Math. Phys. {\bf 1},
516 (1960).
\bibitem{KBI-93} V.E.~Korepin, N.M.~Bogoliubov, and A.G.~Izergin,
{\it Quantum Inverse Scattering Method and Correlation Functions}
(Cambridge University Press, Cambridge, England, 1993).
\bibitem{Gaudin-83}M.~Gaudin, {\it La Fonction d'Onde de Bethe}, (Masson, Paris,
1983).
\bibitem{Korepin-82} V.E.~Korepin, Comm. Math. Phys. {\bf 86}, 391
(1982).
\bibitem{NISolve} \textsf{NISolve} package is available at \\
http://library.wolfram.com/infocenter/MathSource/817/
\bibitem{Wadati} M.~Wadati, J. Phys. Soc. Japan {\bf 71}, 2657
(2002).
\bibitem{Ol} Maxim Olshanii and Vanja Dunjko
Phys. Rev. Lett. {\bf 91}, 090401 (2003).
\end{thebibliography}
\end{document}